\documentclass{icrc}

\usepackage{times}
\usepackage{graphicx} 

\begin{document}

\title{A Project GRAND Study of the GLE of July 14, 2000}
\author[1]{J. Poirier}
\author[1]{C. D'Andrea}
\author[1]{M. Dunford}
\affil[1]{Center for Astrophysics at Notre Dame, Physics Dept., University Of 
Notre Dame, Notre Dame Indiana 46556 USA}

\correspondence{John Poirier (poirier@nd.edu)}

\runninghead{Poirier:GLE}
\firstpage{1}
\pubyear{2001}

\maketitle

\begin{abstract}
The Ground-Level Event (GLE) associated with the X5.7 solar flare of July 14,
2000 is studied by the Project GRAND proportional wire chamber
array.  Results are compared to those obtained by the Climax Neutron Monitor 
experiment
which detects secondary neutrons.  The Climax monitor, located in Climax,
Colorado, is operated by the University of Chicago.  The time of the GLE
signal of the Climax station is examined for a possible coincident signal 
in the data of Project GRAND.  Project GRAND is
an array of 64 proportional wire chamber stations.  The stations are
sensitive to secondary muons with energies greater than 0.1 GeV.  The mean
energy of primaries which produce these ground level muons depends on the
spectral index of the primary spectrum.  For a differential spectral index of 2.4, the most probable 
primary energy is 10 GeV falling rapidly below this energy and falling 
slowly above this energy.
\end{abstract}

\section{Introduction}

Solar flares release a large amount of energy, typically $10^{22}$ to $10^{25}$ 
Joules. Charged particles are also accelerated to high speeds and can 
escape into interplanetary space as solar cosmic rays \citep{co}.  The detection of these 
particles on the Earth at ground level is referred to as a  Ground Level Event (GLE). A 
GLE may be the result of showers of particles produced by the 
interaction between the atmosphere and the high energy primary.  Solar 
flares are classified by their X-ray flux; an X class flare is the 
strongest type of flare and has, for example, a flux in excess of 
$10^{-4}$~W/m$^2$.

An X5.7 solar flare was reported on July 14th, 
2000.  According to X-ray data, the flare lasted from 10:03 UT through
10:43 UT with a peak at 10:24 UT \citep{ips}.  Data from Project GRAND 
\citep{grandweb}, an 
array of proportional wire chambers, is studied and compared to data obtained 
with the Climax Neutron Monitor \citep{cliweb} during the 
time of the flare. 
Project GRAND (41.7$^\circ$N, 
86.2$^\circ$W at 220 m) studies secondary muons and so is more 
likely to see an increase in counting rate that is time coincident with neutron 
monitor data than at the time of the X-ray data.  The Climax Neutron 
Monitor station located in Climax, Colorado (39.4$^\circ$N, 253.8$^\circ$W at 1040 m)
detected a Ground Level Event with onset time of 10:33 UT and peak time of 10:42 UT 
\citep{lopate}.
Their peak flux increased 6\% over background, or a 10 $\sigma$ statistical 
increase.  The data from Climax are shown in Figure 1.

During the peak time of the flare the sun was at 113.6$^\circ \alpha$, 
21.7$^\circ \delta$ \citep{aa2000}; placing the sun at an altitude of 
0.4$^\circ$ and 
an azimuth of 60.7$^\circ$ at Project GRAND at the peak time of the flare.  
The interplanetary magnetic field of the sun intersects the 
Earth at $\sim45^\circ$ angle relative to the Earth-sun line.  The charged particles travel within a 
$45^\circ$ cone relative to this magnetic field \citep{lopate}. This allows 
particles to be detected even though the sun is below the minimum altitude 
 for Project GRAND (27$^\circ$). 
Because the particles still originate from roughly the same general 
direction as the sun, there should be a larger counting rate 
from particles originating from our eastern hemisphere (where the sun was 
located).  Also, because these particles' 
velocity is less than the speed of light and do not follow a straight 
line path, the GLE should be detected at a 
slightly later time than the X-ray event.  

Project GRAND detects ground level muons originating from primaries 
whose energies are dependent on the differential spectral index.  With a 
differential spectral 
index of 2.41, the most probable energy of the primaries is $10$~GeV \citep{jp}; 
a lower index gives a higher primary energy.  Climax detects particles above 
their geomagnetic cutoff energy of 3 GeV. Thus project GRAND is expected 
to see the GLE from the flare slightly earlier than Climax due 
to the higher energies (velocities) of the particles.

\section{Experimental Array}

Project GRAND is an array of 64 proportional wire chamber stations.  Each station is 
composed of four sets of two orthogonal planes with each plane containing 80 detection
 cells of 14 mm width.  The sets of planes are positioned above each other 
 with 200 mm
 of separation between each pair of planes.   This geometrical arrangement allows for 
the measurement of the direction of a
 muon track to within 0.26 degrees, on average, in each of the two projected planes.    
There is also a
 50 mm steel plate above the bottom two planes.  This steel plate 
 and the two additional planes of proportional wire detector planes 
 underneath enable tracks of secondary muons to be distinguished from  
 electron tracks.  
A data rate of $\sim$~2000 muons per second is currently being recorded with the GRAND array.  

Information from each detector station is sent serially to a central 
computor at a rate of 10 MHz.  
The data from the 64 stations are read in parallel into a central data acquisition system 
in 70 microseconds. 
Eight CPU nodes are used in sequence, each in turn receiving data from all 64-stations 
as one event.  The CPU's task is to search the hits from 512 proportional wire planes 
and remember the wire numbers from the eight planes in any station which had 
one and only one hit on each plane of that station.  
Eight nodes operating in sequence perform this 
task with minimal dead time.    
Once a particular node has accumulated data 
 on 900 muons in its buffer memory, the data buffer is written 
as one record on magnetic tape.  
The start time and end time of each record is recorded
along with the wire numbers defining the angles for each of the accumulated 900 muons.  
Radio signals from WWVB in 
Boulder, Colorado, provide time information with precision of a millisecond.  

\balance
\section{Data Analysis}
Project GRAND has a continuous data file starting at 9h UT on July 13 and 
running through 15h UT on July 14, 
allowing easy analysis of the data at the time of the flare as well as providing a
 significant amount of background data. The beginning and end time for each 
 record of 900 muons were
 written to magnetic tape.  The end time of the current record is compared to the end
 time of the previous record from that CPU; the times for each of the
 individual 900 muons are then interpolated evenly within that time interval.  The average
 duration of a record during this time was 3.2 seconds.  

The muon counts are binned in 180 second bins.  Data were examined for the total 
muon count (Figure 2) as a function of time 
 as well as for counts of muons originating only from the eastern hemisphere 
 (Figure 3).  The rms deviation over 
 this time interval is compared to the square root of 
 the number of counts per bin to see if there are additional fluctuations in 
 the data beyond the statistical.  The rms deviation of the muon counts in 
 this interval was 606 compared to the square-root error estimate of 622.   
For muons from only the eastern hemisphere, 
 the rms deviation and square root error were 616 and 434, 
 respectively, suggesting that the data from only the 
 eastern hemisphere perhaps have additional components of 
 variation beyond the statistical.  
The time period of 10.6 to 10.8h is examined using four 180 second bins for a 
possible counting excess.  
The background regions were one hour just before and one hour just after the signal 
region.  This is ten times as much background time as signal time.  The 
muon excess in the signal region was calculated by taking a difference 
between the counts in the signal region minus the average number of 
background counts received in an equal amount of time.   



%


 \begin{figure*}[t] 
 \figbox*{}{}{\includegraphics*[width=11.0cm]{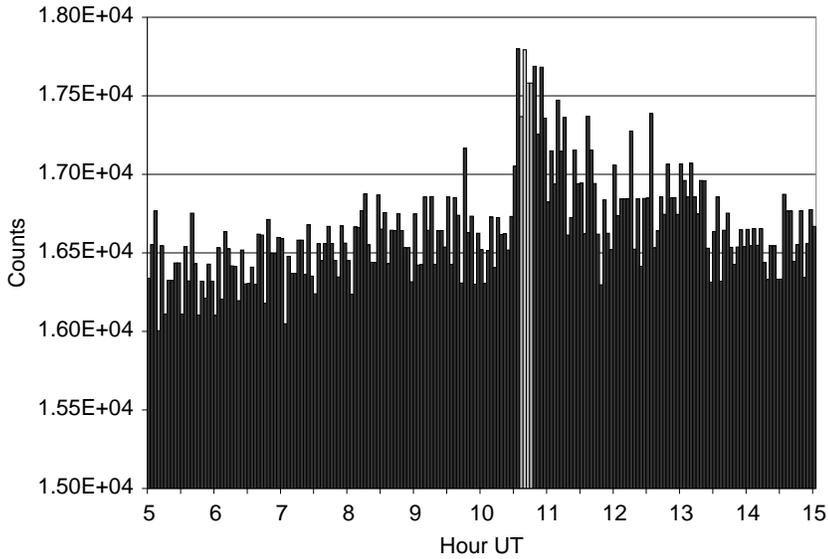}}
 \vspace{-35mm}\caption{Climax Neutron Monitor counting rate 
 (\cite{lopate}).  To guide the eye, the four bins of 
 light shaded area from 10.6 - 10.8 UT corresponds to the same 
light shaded areas of Figures 2 and 3.  The data has been binned in 180 second 
bins for comparison to the data of Project GRAND.}
 \end{figure*}

 \begin{figure*}[t] 
\figbox*{}{}{\includegraphics*[width=11.0cm]{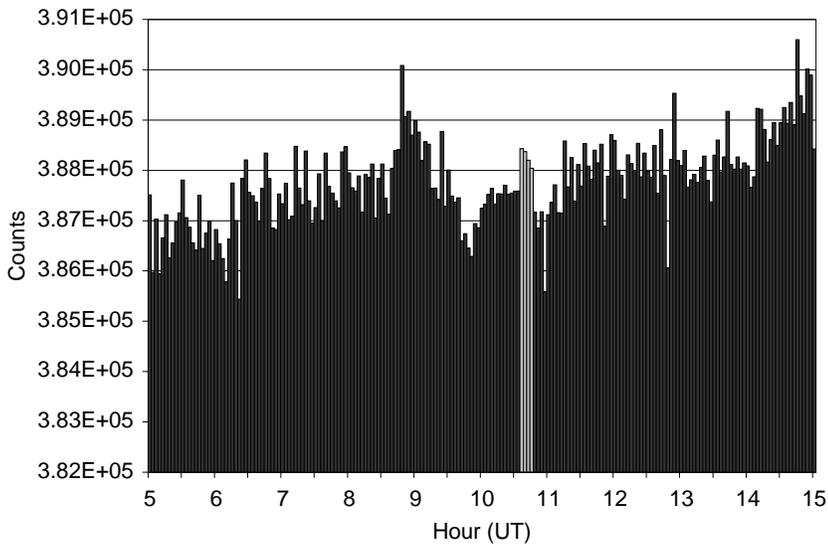}}
 \vspace{-35mm}\caption{Project GRAND's counting rate from east and west hemispheres 
in 180 second bins. 
 The four light bins are from  
	10.6h - 10.8h UT.}
 \end{figure*}

 \begin{figure*}[t] 
 \figbox*{}{}{\includegraphics*[width=11.0cm]{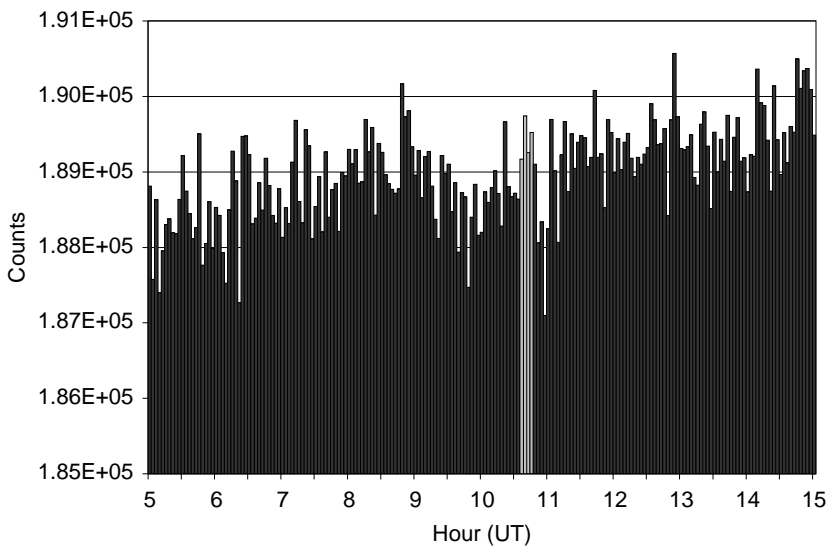}}
 \vspace{-35mm}\caption{Project GRAND's counting rate from the east hemisphere only 
in 180 second bins.  The four light bins are from  
	10.6h - 10.8h UT.}
 \end{figure*}

\section{Conclusions}

Project GRAND detects 
muons which have decayed from pions created predominantly through hadron-air 
interactions.  The Climax Neutron Monitor detects secondary neutrons from interactions between
 the primaries and air.  The Climax GLE has an onset time at 10.55 h and a peak time of 10.70 h. The 
entire period of increased rate is 50 minutes. The peak flux is an 
increase of 6\% (over a ten-sigma statistical increase).  The mean flux 
during the fifty minute period of increased rate is 7$\sigma$ above 
background \citep{lopate}.

Over the interval of 10.6 - 10.8 hrs, using bins of width 180 seconds and a 
background one hour before and one hour after the interval, there is a possible excess above 
background of 3381$\pm$1307 counts, or 2.6 $\sigma$, calculated using 
statistical error (because in this case it is larger than the rms error).  For muons of eastward origin, there 
is a possible excess of 2536$\pm$1295 or 2.0 $\sigma$ (calculated using 
the larger rms error in this case).  If these were interpreted as 
a statistical fluctuation, then they would have a probability of $1.0\%$ 
and $4.6\%$, respectively. 

\begin {acknowledgements}

The authors wish to thank Cliff Lopate for his assistance.  
The Climax Neutron Monitor is operated by the University of Chicago and is supported by National Science Foundation Grant 
ATM-9912341.  Project GRAND is supported through grants from the University 
of Notre Dame and private donations.  

\end {acknowledgements}

\end{document}